\documentstyle[twocolumn,prl,aps,psfig,epsf,floats]{revtex} 

\begin{document}
\title{Heterogeneous Interfacial Failure between Two Elastic Blocks}
\author{G.\ George Batrouni}
\address{Institut Non-Lin\'eaire de Nice, UMR CNRS 6618, 
Universit{\'e} de Nice-Sophia Antipolis, 1361 Route des Lucioles,\\
F--06560 Valbonne, France}
\author{Alex Hansen\footnote{Permanent Address: Department of Physics,
NTNU, N--7491 Trondheim, Norway}}
\address{NORDITA and Niels Bohr Institute, Blegdamsvej 17, DK--2100
Copenhagen {\O}, Denmark}
\author{Jean Schmittbuhl}
\address{Departement de G{\'e}ologie, UMR CNRS 8538,
Ecole Normale Sup{\'e}rieure,
24, rue Lhomond,\\
 F--75231 Paris C{\'e}dex 05, France}
\date{\today}
\maketitle
\begin{abstract} 
We investigate numerically the failure process when two elastic media,
one hard and one soft that have been glued together thus forming a
common interface, are pulled apart.  We present three main results:
(1) The area distribution of simultaneously failing glue (bursts)
follows a power law consistent with the theoretically expected
exponent $2.5$, (2) the maximum load and displacement before
catastrophic failure scale as $L^2$ and $L^0$ respectively, where $L$
is the linear size of the system, and (3) the area distribution of
failed glue regions (clusters) is a power law with exponent $-1.6$
when the system fails catstrophically.
\vspace{0.3cm}

\noindent
PACS number(s): 83.80.Ab, 62.20.Mk, 81.40.Np 
\end{abstract} 
\vskip0.5cm
\section{Introduction}
\label{intro}
The failure of interfaces under stress is a problem that has
obvious important technological relevance.  In addition, from a more
fundamental point of view, this problem exhibits very interesting
features.  It is the aim of this paper to bring out some of these
features by means of a numerical model based on a discretization of
the original problem.

During more than a decade, failure processes in different contexts have 
caught the attention of the physics community.  For a considerable longer
period, the mechanics community has been involved in the study of such
phenomena.  In order to place the present study in its proper context, 
we need to go back to 1926 with the study of Peirce on what today is
known as the {\it global load sharing fiber bundle\/} \cite{p26}.  This consists
of $N$ parallel fibers, each with its own breaking threshold and connected
in such a way that when a fiber fails, the load it was carrying would be
dsitributed equally among all the surviving fibers.  In 1945, Daniels
published a very thorough study of this model, which today forms the starting 
point of any excursion into this field \cite{d45}.  The model has since
these early days been generalized in many directions, one of which consists in
replacing the ``democratic" load-sharing rule by different local ones.
One much studied variant is the {\it local load sharing model,\/} where the
load on the failing fiber is distributed equally among the nearest surviving
fibers \cite{h85}.  The average load--deformation characteristics of the global
load sharing model was calculated by Daniels \cite{d45}.  Corresponding work on
the local load sharing model may be found in  Refs.\ \cite{h85,hp91,dl94}.
There has also been several studies of time-dependent phenomena in
connection with the two variants of the fiber bundle model, see Ref.\
\cite{np01}. (This paper in addition contains a very thorough review of
the literature in this field.) There are also a number of studies ``on the 
market" that may be placed between the two extremes that the global and local 
load sharing models consitute.  Among them, we find the early study by
Newman and Gabrielov \cite{ng91}, who constructed a hierarchically connected
fiber bundle.  Other work on hierarchical fiber bundle models may be found 
in Refs.\ \cite{zhd96,dpcr98,d99}.  

Much work by the physics community has gone into studying network models, of
which the {\it fuse model\/} is the most well known \cite{hr90,b01}.  
This model consists of a network of electrical fuses where their 
burn-out thresholds have been drawn from some probability distribution.  This
model may be regarded as yet another generalization of the fiber bundle model,
however, this time along the axis on which we find chains of fiber bundles 
\cite{sp81}.  Among the several interesting questions that have been studied
in connection with the fuse model, we mention the question of whether the
breakdown process  has the character of a second or first order phase 
transition \cite{zvs97,zrsv99,mgp00,js00}. Central to this question is the
question of the distribution of fuses that burn out simultaneously or --- 
equivalently in the fiber bundle model --- the number of fibers that
fail simultaneously.  This question was first raised and solved analytically
in the context of the global load sharing fiber bundle \cite{hh92} 
and then for the local load sharing model \cite{khh97}.  The same 
question was first studied in connection with the fuse model in Ref.\
\cite{hh94a}.  

The particular problem we study here, elastic interfacial
failure, has been addressed in the literature earlier by Delaplace et
al.\ \cite{d99,drpc99,rdpc99}. The system consists of two elastic
media that have been welded together, thus sharing a common interface.  In
general, the media can have different elastic constants.  However, for the
sake of simplicity and without loss of generality, we assume one of the
media to be infinitely stiff while the other is elastic.  We can view this
simplification as an effective representation of the original system since
it does not change the physics.  Furthermore, we assume that the ``soft''
medium is uniform with respect to its elastic properties; it has the same
elastic properties everywhere.  However, the local {\it strength\/} of the
glue --- defined as the maximum local load it may sustain without failing
--- varies from point to point along the interface.  This is the source of
disorder in the system.  In real systems this disorder would typically be
correlated.  In this first attack on the problem, we assume the disorder to
be uncorrelated. Our main interest here is to understand how correlations
develop due to the failure process. In Section \ref{model} we describe the
numerical model in detail.

The two joined media are subjected to a progressive uniform load
perpendicular to the glue interface.  Local failures will develop in
the interface which changes the stress field on the remaining intact
interface.  These changes in the stress field will compete with the
local strength of the glue to determine where the next failure
happens.  Sometimes, a local failure will occur due to the glue being
particularly weak at that point on the interface, other times failure
will occur due to enhancements in the local stress field.  This
competition leads to the development of spatial correlations both in
the stress field and in the failure patterns, and in the failure
process itself.

The two media can be pulled apart by controlling (fixing) either the
applied force or the {\it displacement}. The displacement is defined
as the change in the distance between two points, one in each medium,
positioned far from the glued interface. Clearly, the line connecting
these points is perpendicular to the average position of the
interface. In our case, the pulling is accomplished by controlling the
displacement. As the displacement is increased very slowly, glued
points will fail. Sometimes the failed regions are very small, other
times, the failed region is larger. Such events, when a large area
fails ``intantaneously'' compared to the time scale at which the
displacement is changed, are called bursts. One of the quantities of
interest to us is the {\it burst distribution\/} \cite{hh92} as the
failure process evolves.  We show in Section
\ref{bursts} that this distribution follows a power law.

In Section \ref{scaling}, we investigate the scaling properties of the
load and displacement of the system at the point when the failure
process becomes unstable.  This is the point at which any further
increase of either load or displacement will lead to a catastrophic
burst where all remaining glue fails.  This point defines the {\it
strength\/} of the interface, and the question we pose is how this
scales with the system size.

We then investigate the geometrical properties of the failed regions
at the point when catastrophic failure occurs in Section \ref{damage}.
We find that the area distribution of the failed regions follow a
power law.

We present our conclusions and outlook for further work in Section
\ref{conc}.

\section{Model}
\label{model}

The system described in the Introduction is continuous: Two media, one
elastic, the other infinitely stiff, are glued together thus forming a
common interface.  In order to treat this problem numerically, the
continuum problem is replaced by a discrete one.  We use a discrete
model for this.  In Section \ref{desc} we describe the discrete model,
while the numerical algorithms are discussed in Section
\ref{numerics}.

\subsection{Description of Model}
\label{desc}

We discretize the glued interface by replacing it with two
two-dimensional square $L\times L$ lattices with periodic boundary
conditions. The lower one represents the hard, stiff surface and the
upper one the elastic surface. The nodes of the two lattices are
matched ({\it i.e.\/} there is no relative lateral displacement). The
glue is modelled by springs connecting opposing nodes in the two
lattices. These harmonic springs all have the same spring constant
(set to unity) but breaking thresholds randomly drawn from a uniform
distribution between zero and one. As the two glued media are
separated, the forces carried by the springs increases. When the force
carried by a spring reaches its breaking threshold, it breaks
irreversibly and the forces redistribute.  The springs are thus broken
one by one until the two media are no longer in mechanical contact. As
this process is proceeding, the elastic body is of course deforming to
accomodate the changes in the forces acting on it.

The equations governing the system are as follows. The force, $f_i$,
carried by the $i$th spring is given by Hooke's law,
\begin{equation}
\label{M1}
f_i = -k(u_i-D)\;,
\end{equation}
where $k$ is the spring constant, $D$ is the displacement of the hard
medium, and $u_i$ is the deformation of the elastic medium at site
$i$. All unbroken springs have $k=1$ while a broken spring has
$k=0$. The quantity $(u_i-D)$ is, therefore, the length spring $i$ was
stretched. In addition, a force applied at a point on an elastic
surface will deform this surface over a region whose extent depends on
its elastic properties. This is described by the coupled system of
equations,
\begin{equation}
\label{M2}
u_i=\sum_{j} G_{i,j} f_j\;,
\end{equation}
where the elastic Green function, $G_{i,j}$ is given by\cite{ll58,j85}
\begin{equation}
\label{M3}
G_{i,j} = \frac{1-s^2}{\pi E a^2}\ \int_{-a/2}^{+a/2}\
\frac{dx'\ dy'}{|(x-x',y-y')|}\;.
\end{equation}
In this equation, $s$ is the Poisson ratio, $e$ the elastic constant,
and $|{\vec i}-{\vec j}|$ the distance between sites $i$ and $j$. The
indices $i$ and $j$ run over all $L^2$ sites.  Strictly speaking, this 
Green function applies for a medium occupying the infinite half
space. However, with a judicious choice of elastic constants, we may
use it for a finite medium if its range is small compared to $L$, the
size of the system.  

By combining equations (\ref{M1}) and (\ref{M2}), we obtain
\begin{equation}
\label{M4}
({\bf I} +{\bf K G}) {\vec f} = {\bf K} {\vec D}\;,
\end{equation}
where we are using matrix-vector notation. ${\bf I}$ is the $L^2\times
L^2$ identity matrix, and ${\bf G}$ is the Green function represented
as an $L^2\times L^2$ dense matrix. The constant vector ${\vec D}$ is
$L^2$ dimensional. The {\it diagonal\/} matrix ${\bf K}$ is also
$L^2\times L^2$. Its matrix elements are either 1, for unbroken
springs, or 0 for broken ones. Of course ${\bf K}$ and ${\bf G}$ do
not commute (except initially when there are no broken springs).

Once equation (\ref{M4}) is solved for the force, ${\vec f}$, equation
(\ref{M2}) easily yields the deformations of the elastic surface.

\subsection{Numerical Method: Fourier Acceleration}
\label{numerics}

Equation (\ref{M4}) is of the familiar form ${\bf A}{\vec x}={\vec
b}$.  Since the Green function connects all nodes to all other nodes,
the $L^2\times L^2$ matrix ${\bf A}$ is dense which puts severe limits
on the size of the system that may be studied.  There are direct, time
consuming methods to deal with such matrices, see Ref. \cite{ptvf92}.
However, as we shall see, this problem may be circumvented and much
more efficient methods may be employed such as the Conjugate Gradient
algorithm (CG) \cite{ptvf92,bh88}.

The simulation proceeds as follows: We start with all springs present,
each with its randomly drawn breakdown threshold. The two media are
then pulled apart, the forces calculated using CG, and the spring
which is the nearest to its threshold is broken, {\it i.e.\/} the
matrix element corresponding to it in the matrix ${\bf K}$ is
zeroed. Then the new forces are calculated, a new spring broken and so
on till all springs have been broken and the media separated.

However, there are two problems that render the simulation of large
systems extremely difficult. The first is that since ${\bf G}$ is
$L^2\times L^2$ {\it dense\/} matrix, the number of operations per CG
iteration scales like $L^4$. Even more serious is the fact that as the
system evolves and springs are broken, the matrix $({\bf I}+k{\bf G})$
becomes very ill-conditioned.

\begin{figure}
\epsfxsize=3.3in
\epsfysize=2.25in
\epsffile{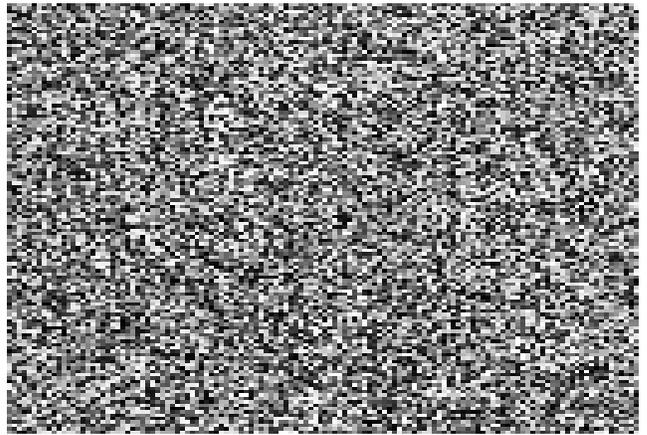}
\caption{Each elementary square represents a bond.  The color scheme 
indicates when in the failure process a given bond failed, the
lighter, the earlier.  The lattice was $128\times 128$ with an elastic
constant $e=10$.}
\label{fig1}
\end{figure}

To overcome the problematic $L^4$ scaling of the algorithm we note
that the Green function is diagonal in Fourier space. Consequently,
doing matrix-vector multiplications using FFTs the scaling is much
improved and goes like $L^2 \ln(L)$.  Symbolically, this can be
expressed as follow:
\begin{equation}
\label{M5}
({\bf I} + {\bf K} {\bf F^{-1}F}{\bf G}){\bf F^{-1}F} {\vec f} = {\bf
K} {\vec D}\;,
\end{equation}
where ${\bf F}$ is the FFT operator and ${\bf F^{-1}}$ its inverse
(${\bf F^{-1}F}=1$). Since ${\bf I}$ and ${\bf K}$ are diagonal,
operations involving them are performed in real space. With this
formulation, the number of operations/iteration in the CG algorithm
now scales like $L^2\ln(L)$.

To overcome the runaway behavior due to the ill-conditioning we need
to precondition the matrix \cite{bh88}. This means that instead of
solving equation (\ref{M5}), we solve the equivalent problem
\begin{equation}
\label{M6}
{\bf Q}({\bf I} + {\bf K} {\bf F^{-1}F}{\bf G}){\bf F^{-1}F} {\vec f} 
= {\bf Q} {\bf K} {\vec D}\;,
\end{equation}
where we simply multiplied both sides by the arbitrary, positive
definite preconditioning matrix ${\bf Q}$. Clearly, the ideal choice
is ${\bf Q_{0}}=({\bf I} + {\bf K}{\bf G})^{-1}$ which would always
solve the problem in one iteration. Since this is not possible in
general, we look for a form for ${\bf Q}$ which satisfies the
following two conditions: (1) As close as possible to ${\bf Q_{0}}$,
and (2) fast to calculate. The choice of a good ${\bf Q}$ is further
complicated by the fact that as the system evolves and springs are
broken, corresponding matrix elements of ${\bf K}$ are set to
zero. So, the matrix $({\bf I} + {\bf K}{\bf G})$ evolves from the
initial form $({\bf I} + {\bf G})$ to the final one ${\bf I}$. We
were not able to find a fixed ${\bf Q}$ that worked throughout the
breaking process.

We therefore chose the form
\begin{equation}
\label{M7}
{\bf Q}={\bf I}-({\bf K}{\bf G})+({\bf K}{\bf G})({\bf K}{\bf G})
-({\bf K}{\bf G})({\bf K}{\bf G})({\bf K}{\bf G})+ ...
\end{equation}
which is nothing but the Taylor series expansion of ${\bf Q_{0}}=({\bf
I} + {\bf K}{\bf G})^{-1}$. For best performance, the number of terms
kept in the expansion is left as a parameter since it depends on the
physical parameters of the system. It is important to emphasize the
following points. (a) As springs are broken, the preconditioning
matrix evolves with the ill-conditioned matrix and, therefore, remains
a good approximation of its inverse throughout the breaking
process. (b) All matrix multiplications involving ${\bf G}$ are done
using FFTs. (c) The calculation of ${\bf Q}$ can be easily organized
so that it scales like $n L^2 \ln(L)$ where $n$ is the number of term
kept in the Taylor expansion, equation (\ref{M7}).

We therefore have a stable accelerated algorithm which scales
essentially as the volume of the system. For example, for a $128\times128$
system, and taking $n=5$, the CG algorithm always converged in four
or five iterations with the prescribed precision of $10^{-12}$.

\section{Results}
\label{results}

\begin{figure}
\epsfxsize=3.3in
\epsfysize=2.25in
\psfig{file=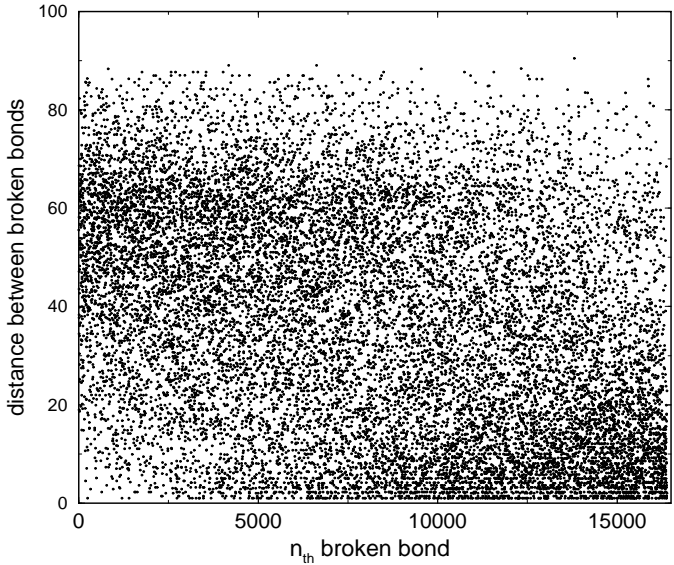,height=3.0in,width=3.5in,angle=-90}
\caption{Distance between succesively broken bonds. The lattice 
was $128\times 128$ with an elastic constant $e=10$.}
\label{fig2}
\end{figure}

\begin{figure}
\epsfxsize=3.3in
\epsfysize=2.25in
\psfig{file=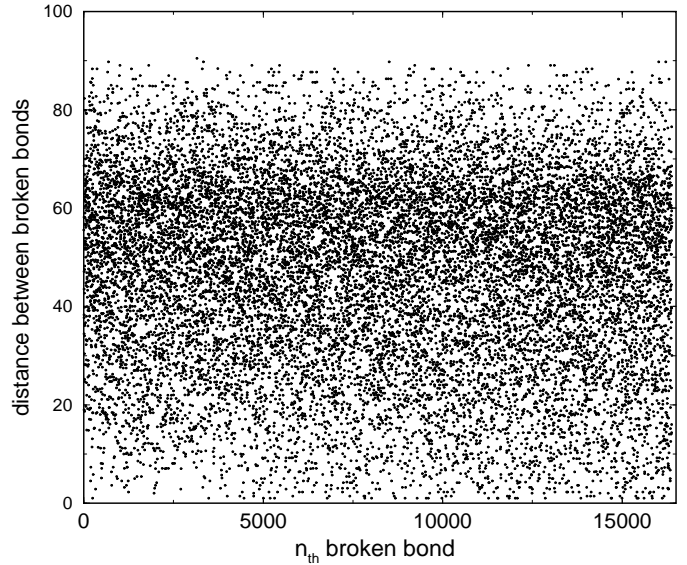,height=3.0in,width=3.5in,angle=-90}
\caption{Distance between succesively broken bonds.
The lattice was $128\times 128$ with an elastic constant $e=100$.
\label{fig3}}
\end{figure}

We now present the results of our numerical simulations.  We show in
Fig.\ \ref{fig1} a representation of the failure process.  Each
elementary square represents a spring (a bond), and the gray scale
indicates when a particular spring failed: The darker the color, the
earlier the failure.  In this particular case, the elastic constant
$e$ was set to 10.  There are no apparent spatial correlations between
the failing bonds in this figure.  However, we show in Fig.\
\ref{fig2} the distance between successively failing bonds for the
same disorder realization of Fig.\ \ref{fig1}.  We see clearly in this
figure that early in the process there is no localization effect:
Bonds tend to break far apart, the location being determined by the
strength of bonds, {\it i.e.} early failure is disorder driven.
However, halfway into the breakdown process, localization clearly
develops.  In Fig.\ \ref{fig3}, we show the corresponding plot for
$e=100$. In this case localization never develops for this size system
and destribution of thresholds.

\begin{figure}
\epsfxsize=3.3in
\epsfysize=2.25in
\epsffile{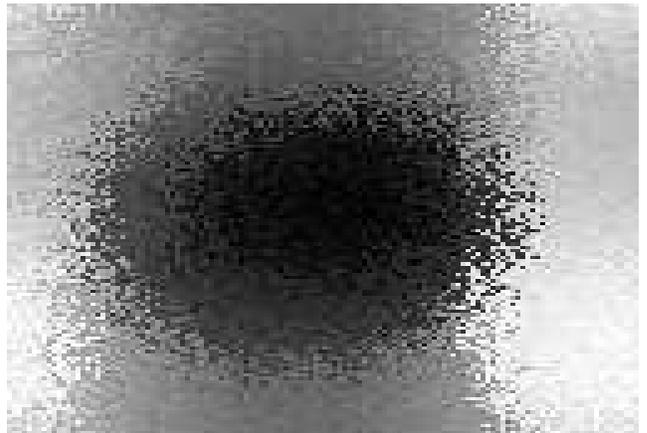}
\caption{Same as in Fig.\ \ref{fig1} but with a narrow uniform
threshold distribution in the interval $[9.5,10.5]$.}
\label{fig4}
\end{figure}

On the other hand, if the threshold distribution is much narrower than
$[0,1]$ used above, localization can develop early. For example, we
show in Fig.\ \ref{fig4} the fracture graymap (like Fig.\ \ref{fig1})
for a uniform threshold distribution in the interval $[9.5,10.5]$. We
clearly see the fracture starting towards the center of the figure and
spreading out in a spiral till finally the symmetry is broken and the
system ruptures along one of the lattice directions.

\begin{figure}
\epsfxsize=3.3in
\epsfysize=2.25in
\psfig{file=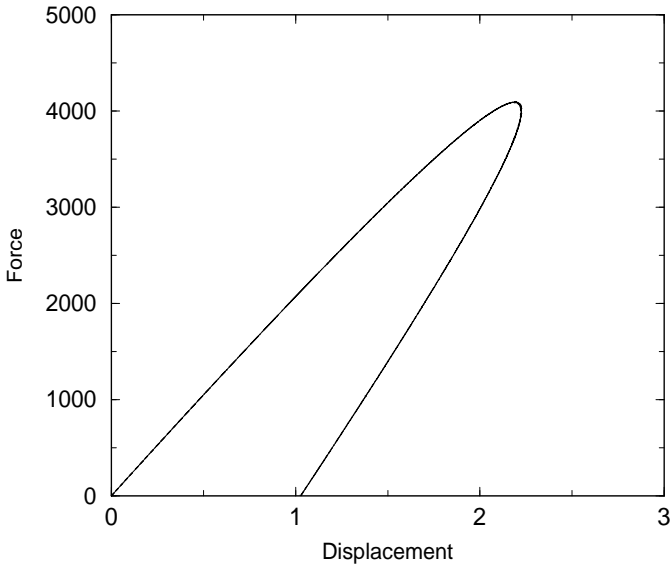,height=3.0in,width=3.5in,angle=-90}
\caption{Force-displacement curve, $128\times 128$ systems with $e=10$.
\label{fig5}}
\end{figure}

\begin{figure}
\epsfxsize=3.3in
\epsfysize=2.25in
\psfig{file=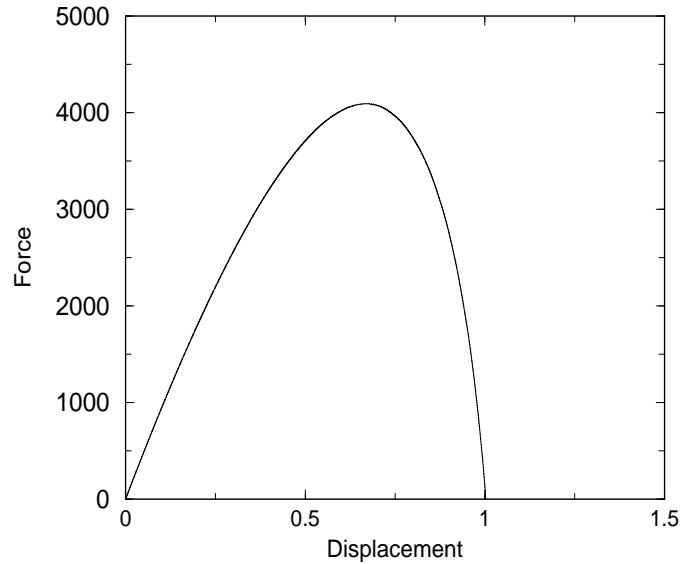,height=3.0in,width=3.5in,angle=-90}
\caption{Force-displacement curve, $128\times 128$ systems with $e=100$.
\label{fig6}}
\end{figure}

Fig.\ \ref{fig5} shows the force-displacement curve for a system with
elastic constant $e=10$.  Whether we control the applied force, $F$,
or the displacement, $D$, the system will eventually suffer
catastrophic collapse.  However, this is not so when $e=100$ as shown
in Fig.\ \ref{fig6}.  In this case, only controlling force will lead
to catastrophic failure.  In the limit when $e\to\infty$, the model
becomes the global load-sharing fiber bundle model \cite{p26,d45},
where $F=(1-D)D$.  In this limit there are no spatial correlations and
the force instability is due to the the decreasing total elastic
constant of the system making the force on each surviving bond
increase faster than the typical spread of threshold values. No such
effect exists when controlling displacement $D$.  However, when the
elastic constant, $e$, is small, spatial correlations in the form of
localization do develop, and these are responsible for the diplacement
instability which is seen in Fig.\ \ref{fig5}. In other words, the
localization clearly visible in Fig.\ \ref{fig2} starts to develop
when the system is near the peak of its force-displacement curve, and
dominates when the system is on the negative slope branch of that
curve.

\subsection{Burst Distribution}
\label{bursts}

We now turn to the study of the burst distribution.  We define the
size of a burst, $\Delta$, in our model as the number of bonds that
fail simultaneously while the force $F$ is held constant.  In the
global load-sharing fiber bundle model it has been shown that the
burst distribution is given by \cite{hh92}
\begin{equation}
\label{fiberbundle}
N(\Delta,D)=\frac{1}{\Delta^\tau}\ n\left(\Delta^\sigma (x-x_c)\right)\;
\end{equation}
where $x_c$ is the damage, i.e. the density of broken bonds, at which
the model fails catastrophically, $n$ is a crossover function which
approaches a constant when the argument approaches zero, and which
falls off as $\exp(-y^2)$ as the argument $y$ is large.  Furthermore,
\begin{equation}
\label{burstexp}
\tau=\frac{5}{2}\qquad {\rm and} \qquad \sigma=\frac{1}{2}
\end{equation}
independent of the threshold distribution.

\begin{figure}
\epsfxsize=3.3in
\epsfysize=2.25in
\psfig{file=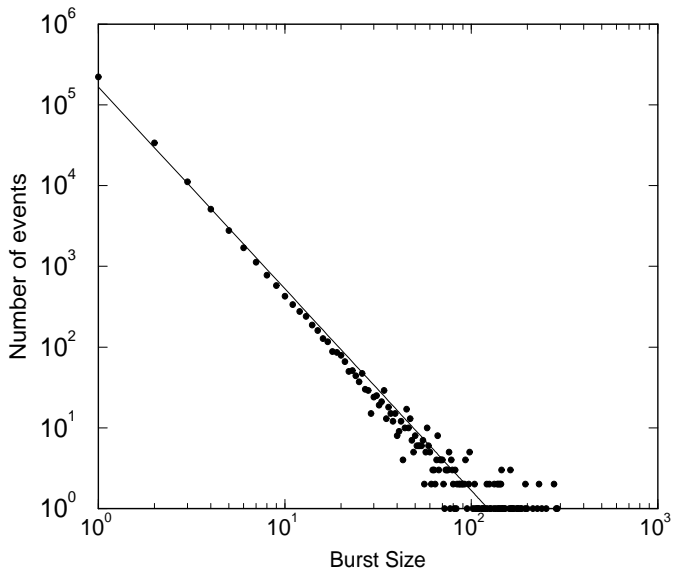,height=3.0in,width=3.5in,angle=-90}
\caption{Burst distribution for $128\times 128$, $e=10$.  The slope of 
the straight line is $-2.5$.
\label{fig7}}
\end{figure}

\begin{figure}
\epsfxsize=3.3in
\epsfysize=2.25in
\psfig{file=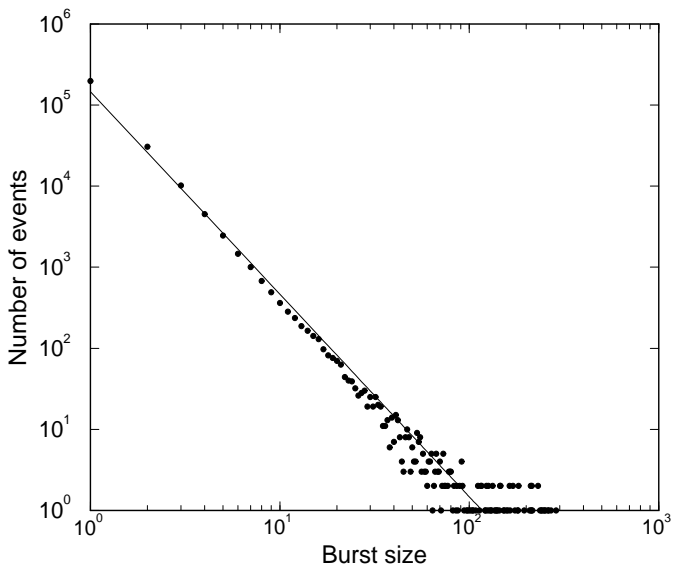,height=3.0in,width=3.5in,angle=-90}
\caption{Burst distribution for $128\times 128$, $e=100$.  The slope of 
the straight line is $-2.5$.
\label{fig8}}
\end{figure}

We show in Figs.\ \ref{fig7} and \ref{fig8} the burst distribution for
$e=10$ and 100. In both cases we find that the burst distribution
follows a power law with an exponent $\tau=-2.6\pm0.1$.  We may argue
that the exponent is the same as the one found in the global loading
fiber bundle, Eq.\ (\ref{burstexp}), $\tau=5/2$ in the following way.
The characteristics, $F=F(x)$ must have a quadratic maximum somewhere.
For $e=100$, such a maximum exists in the middle of failure process as
seen in Fig.\ \ref{fig8}, whereas for $e=10$, the system only
approaches such a maximum near total failure, see Fig.\ \ref{fig7}.
Assuming that the fluctuations about the average characteristics are
brownian --- which can be shown analytically in the limit $e\to\infty$
\cite{pt73,ds85,d89} --- near the maximum the probability to find a
burst of size $\Delta$ is proportional to $\Delta^{-3/2}\exp[-\Delta
(x-x_c)^2]$.  This result comes from a mapping onto the {\it Gambler's
ruin problem\/} \cite{hh94b}.  Furthermore, in order to guarantee that
the burst is not a burst within an even larger burst, the starting
point of the burst must be the highest point on the characteristics
that has occured so far in the failure process.  This condition may
also be mapped onto the Gambler's ruin problem, and leads to an extra
factor $(x-x_c)$ in the probability for a burst to occur.  The
probability to find a burst of size $\Delta$ throughout the failure
process is then the integral over $x$ as $x$ approaches $x_c$,
$\int^{x_c} dx (x-x_c) \Delta^{-3/2}\exp[-\Delta (x-x_c)^2]$ which is
proportional to $\Delta^{-5/2}$.

As can be seen in Figs.\ \ref{fig7} and \ref{fig8}, the numerical data
are consistent with the expected value $\tau=5/2$.

\subsection{Strength Scaling}
\label{scaling}

\begin{figure}
\epsfxsize=3.3in
\epsfysize=2.25in
\psfig{file=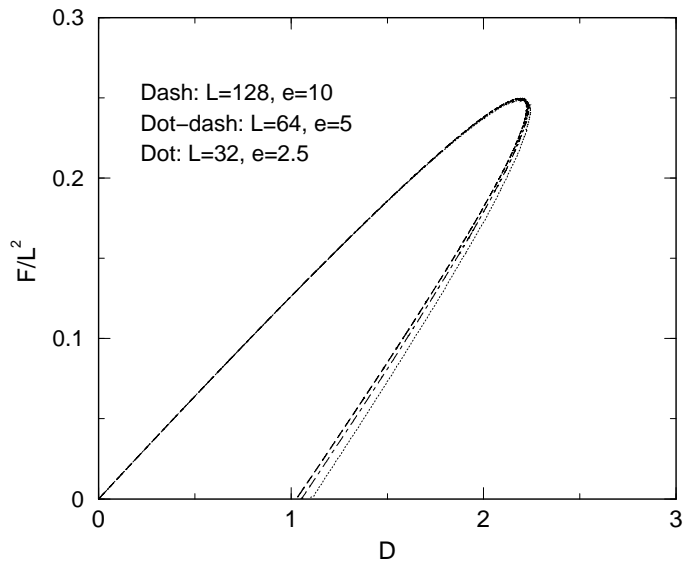,height=3.0in,width=3.5in,angle=-90}
\caption{Scaling of the failure characteristics for systems with
$L=128 (e=10), 64 (e=5)$ and $32 (e=2.5)$ using the reduced variables
$F/L^{2.0}$ and $D/L^{0.0}$.}
\label{fig9}
\end{figure}

The load-displacement curves for different system sizes $L$ coincide
when we use the reduced variables $F/L^a$ and $D/L^b$ where $a=2.0$
and $b=0.0$, as seen in Fig.\ \ref{fig9}.  We expect the exponent
$a=2$ since $F/L^2$ is the normal stress on the surface.  In the case
of an infinitely stiff system, we expect $b=0$.  The elastic system
studied here behaves in the same way as long as the elastic constant,
$e$, is also scaled with $L$. For example for $L=128$ we took $e=10$,
for $L=64$ $e$ should take half that value in order to reproduce the
physics. This is easy to understand considering the dependence of the
Green fucntion, Eq.\ \ref{M3}, on the elastic constant.

\subsection{Spatial Damage Distribution at Failure}
\label{damage}

\begin{figure}
\epsfxsize=3.3in
\epsfysize=2.25in
\psfig{file=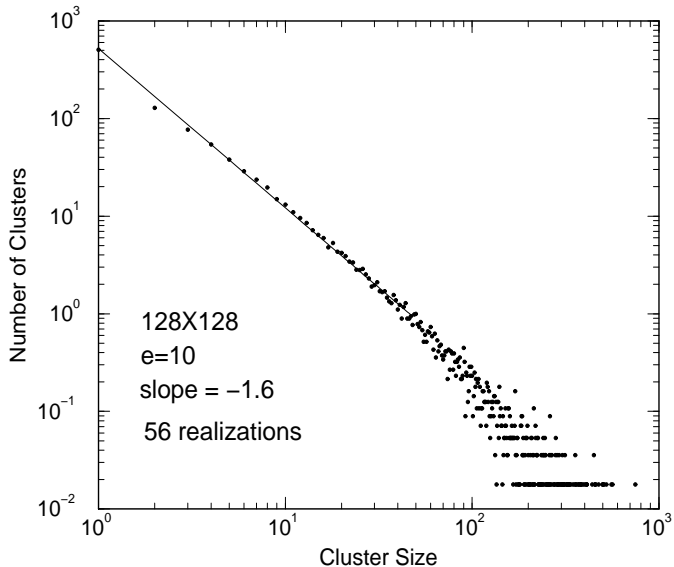,height=3.0in,width=3.5in,angle=-90}
\caption{Area distribution of zones where glue has failed for systems of
size $128\times 128$ and elastic constant $e=10$.  The straight line
is a least square fit and indicates a power law with exponent $-1.6$.
$e=10$
\label{fig10}}
\end{figure}

As the failure process proceeds, there is an increasing competition
between local failure due to stress enhancement and local failure due
to local weakness of material.  As we saw above, when we control the
displacement, $D$, and $e$ is sufficiently small (for example $e=10$),
catastrophic failure eventually occurs due to localization. The onset
of this localization, {\it i.e.} the catastrophic regime, occurs when
the two mechanisms are equally important. One may suspect that
criticality due to self organization \cite{btw87} occurs at this
point.  In order to test whether this is the case, we have measured
the size distribution of broken bond clusters at the point when $D$
reaches its maximum point on the $F-D$ characteristics, {\it i.e.} the
onset of localization and catastrophic failure.  The analysis was
performed using a Hoshen-Kopelman algorithm \cite{sa94}.  We show the
result in Fig.\ \ref{fig10}, for 56 disorder realizations, $L=128$ and
$e=10$. The result is consistent with a power law distribution with
exponent $-1.6$, and consequently with self organization.

\section{Conclusion}
\label{conc}

We have studied numerically the failure of the glued interface between
an elastic and an infinitely stiff blocks of material.  To this end we
have developed a new, stable and accelerated algorithm which scales
$L^2\ln(L)$ which enabled us to study much bigger systems than
previously possible.

Our main physical results are: (1) the distribution of simultaneously
failing glue (bursts) is a power law with exponent $-2.6\pm0.1$, which
is consistent with the burst distribution found in the global load
sharing fiber bundle problem.  (2) The point of catastrophic failure
scales as $L^2$ in force and $L^0$ in displacement.  (3) The area
distribution of failed regions (clusters) at the onset of catastrophic
failure when displacement is the control parameter is consistent with
a power law with an exponent equal to $-1.6$.  This hints at self
organization.

In addition, we saw that for large $e$, e.g. $e=100$, the system does
not suffer catastrophic failure, and there is no localization. On the
other haand, smaller values of $e$, e.g. $e=10$, resulted in
catastrophic failure due to localization. By doing the simulations for
various values of $e$ we estimate that failure due to localization
starts to occur for $e\sim 35-40$. As we will see below, these values
of $e$ obtained for $128 \times 128$ systems should be scaled
appropriately when the size of the system is changed.

The disorder in our system was uncorrelated. As mentioned above, it is
realistic to introduce correlations as exist for example in fracture 
surfaces. This can be done by generating spring breaking thresholds 
that have the desired correlations. Furthermore, we have used a flat
distribution for the disorder. One can use other distributions,
e.g. $r^{\alpha}$, where $r$ is a uniformly distributed random number
and $\alpha$ an exponent that can be negative. It is known from random
fuse models of fracture that the breakdown process depends on the
value of $\alpha$. It is not clear how these issues will modify our
current results. This work is in progress.

Another work in progress is to study the propagating fracture front
when the above glued media are ripped apart by pulling only on one
side of the $L\times L$ square system. As before the breaking
thresholds can be correlated or uncorrelated. These results will also
be compared with the results of experiment currently underway. 

Finally, we have chosen to introduce disorder into the breaking
thresholds of the springs. However, we can just as easily introduce it
into the spring constants themselves, again with or without
correlations.

\acknowledgments

We thank F.\ A.\ Oliveira and H.\ Nazareno and the ICCMP of the
Universidade de Bras{\'\i}lia for warm hospitality and support during
the initial phases of this work. This work was partially funded by the
CNRS PICS contract $\#753$ and the Norwegian research council, NFR.
We also thank NORDITA for its hospitality and further support.  A.\
H.\ thanks the Niels Bohr Institute for their support and hospitality
during his sabbatical year.



\begin{thebibliography}{10}

\bibitem{p26} F.\ T.\ Peirce,  J.\ Text.\ Ind.\ {\bf 17}, 355 (1926).

\bibitem{d45} H.\ E.\ Daniels,  Proc.\ Roy.\ Soc.\ London {\bf A183},
405 (1945).

\bibitem{h85} D.\ G.\ Harlow, Proc.\ Roy.\ Soc.\ Lond.\ Ser.\ A
{\bf 397}, 211 (1985).

\bibitem{hp91} D.\ G.\ Harlow and S.\ L.\ Phoenix,  J.\ Mech.\ Phys.\
Solids {\bf 39}, 173 (1991).

\bibitem{dl94} P.\ M.\ Duxbury and P.\ M.\ Leath,  Phys.\ Rev.\ B {\bf 49},
12676 (1994).

\bibitem{np01} W.\ I.\ Newman and S.\ L.\ Phoenix, Phys.\ Rev.\ E {\bf 63},
021507 (2001).

\bibitem{ng91} W.\ I.\ Newman and A.\ M.\ Gabrielov, Int.\ J.\ Fracture
{\bf 50}, 1 (1991).

\bibitem{zhd96} S.\ -d.\ Zhang, Z.\ -q.\ Huang and E.\ -j.\ Ding,
Phys.\ Rev.\ E {\bf 54}, 3314 (1996).

\bibitem{dpcr98} A.\ Delaplace, G.\ Pijaudier-Cabot and S.\ Roux,
J.\ Physique IV, {\bf 8}, 103 (1998).

\bibitem{d99} A.\ Delaplace, {\sl Analyse Statistique de la Localization 
dans les Mat{\'e}riaux H{\'e}t{\'e}rog{\`e}nes Quasi-Fragiles,\/} 
Doctoral Thesis (Ecole Normale Sup{\'e}rieure,
Cachan, 1999).

\bibitem{hr90} H.\ J.\ Herrmann and S.\ Roux, {\sl Statistical Models for the
Fracture of Disordered Media\/} (North-Holland, Amsterdam, 1990).

\bibitem{b01} A. Hansen in {\sl Physical Aspects of Fracture,\/} ed.\ by
E.\ Bouchaud (Kluwer Academic Publishers, Dordrecht, 2001).

\bibitem{sp81} R.\ L.\ Smith and S.\ L.\ Phoenix, ASME J.\ Appl.\
Mech.\ {\bf 48}, 75 (1981).

\bibitem{zvs97} S.\ Zapperi, A.\ Vespignani and H.\ E.\ Stanley,
Nature {\bf 388}, 658 (1997).

\bibitem{zrsv99} S.\ Zapperi, P.\ Ray, H.\ E.\ Stanley and A.\ Vespignani,
Phys.\ Rev.\ E {\bf 59}, 5049 (1999).

\bibitem{mgp00} Y.\ Moreno, J.\ B.\ G{\'o}mez and A.\ F.\ Pacheo, Phys.\ 
Rev.\ Lett.\ {\bf 85}, 2865 (2000). 

\bibitem{js00} A.\ Johansen and .\ Sornette, Eur.\ Phys.\ J.\ B {\bf 18},
163 (2000).

\bibitem{hh92} P.\ C.\ Hemmer and A.\ Hansen, ASME J.\ Appl.\ Mech.\ 
{\bf 59}, 909 (1992).

\bibitem{khh97} M.\ Kloster, A.\ Hansen and P.\ C.\ Hemmer, Phys.\ Rev.\
E {\bf 56}, 2615 (1997).

\bibitem{hh94a} A.\ Hansen and P. C.\ Hemmer, Phys.\ Lett.\ A {\bf 184}, 
394 (1994).

\bibitem{drpc99} A.\ Delaplace, S.\ Roux and G.\ Pijaudier-Cabot,
Int.\ J.\ Sol.\ Struct.\ {\bf 36}, 1403 (1999).  

\bibitem{rdpc99} S.\ Roux, A.\ Delaplace and G.\ Pijaudier-Cabot,
Physica A, {\bf 270}, 35 (1999).

\bibitem{ll58} L.\ Landau and E.\ M.\ Litshitz, {\sl Theory of Elasticity\/}
(Clarendon Press, Oxford, 1958).

\bibitem{j85} K.\ L.\ Johnson, {\sl Contact Mechanics\/} (Cambridge University
Press, Cambridge, 1985).

\bibitem{ptvf92} W.\ H.\ Press, S.\ A.\ Teukolsky, W.\ T.\ Vetterling and
B.\ P.\ Flannery, {\sl Numerical Recipes in Fortran 77: The Art of
Scientific Computing\/} (Cambridge University Press, Cambridge, 1992).

\bibitem{bh88} G.\ G.\ Batrouni and A.\ Hansen, J.\ Stat.\ Phys.\ {\bf 52},
747 (1988).

\bibitem{pt73} S.\ L.\ Phoenix and H.\ M.\ Taylor, Adv.\ Appl.\ Prob.\ 
{\bf 5}, 200 (1973).

\bibitem{ds85} H.\ E.\ Daniels and H.\ T.\ R.\ Skyrme, Adv.\ Appl.\ Prob.\
{\bf 17}, 85 (1985).

\bibitem{d89} H.\ E.\ Daniels, Adv.\ Appl.\ Prob.\ {\bf 21}, 315 (1989).

\bibitem{hh94b} A.\ Hansen and P.\ C.\ Hemmer, Trends in Statistical Physics,
{\bf 1}, 213 (1994).

\bibitem{btw87} P.\ Bak, C.\ Tang and K.\ Wiesenfeld, Phys.\ Rev.\ Lett.\
{\bf 59}, 381 (1987).

\bibitem{sa94} D.\ Stauffer and A.\ Aharony, {\sl Introduction to Percolation
Theory\/} (Taylor and Francis, London, 1994).

\end{thebibliography}
\end{document}